\documentstyle[pra,aps,preprint]{revtex}
\begin{document}
\draft
\title{Edge and Bulk of the Fractional Quantum Hall Liquids}
\author{Naoto Nagaosa}
\address{Department of Applied Physics, University of Tokyo,
Bunkyo-ku, Tokyo 113, Japan}

\author{Mahito Kohmoto}
\address{Institute for Solid State Physics,  University of Tokyo,
Roppongi, Minato-ku, Tokyo 106, Japan}

\date{\today}
\maketitle
\begin{abstract}
  An effective Chern-Simons theory for the Abelian quantum Hall states
with edges is proposed to study the edge and bulk properties in a
unified fashion. We impose a condition that the currents do not flow
outside the sample.  With this boundary condition, the action remains gauge
invariant and the edge modes are naturally derived.
We find that the integer coupling matrix $K$ should satisfy the condition
$\sum_I(K^{-1})_{IJ} = \nu/m$ ($\nu$: filling of Landau levels,
$m$: the number of gauge fields ) for the quantum Hall liquids.
Then the Hall conductance is always quantized irrespective of
the detailed dynamics or the randomness at the edge.
\end{abstract}


 Recently the chiral Tomonaga-Luttinger (TL) liquid \cite{r1}
realized at the edges
of fractional quantum Hall liquid (FQHL) \cite{r2} has been proposed to be
a promising ideal one-dimensional system among the ones which have ever
been studied. In the quantum wire with zero magnetic field, the
randomness causes backward scatterings and hence the localization.
In contrast, it can be avoided in the QHL due to the following
reason \cite{r3}.
The probability of the
backward scatterings is proportional to the overlap of the wavefunctions of
the edge modes with the opposite chiralities.
When the chiralities of the edge modes in one edge of the sample are
all the same and the other edge is spatially well separated,
this probability is exponentially small and can be neglected.
Hence the Hall conductance of the edge is robust against the randomness
and is quantized, which is the explanation of the (F)QHE in terms of the
edge picture.
When one intentionally introduces
the backward scatterings between the two edges by making a point contact,
the system is expected to be described as the TL model with
potential barriers \cite{r4,r5}. This idea \cite{r5} beautifully
explains the recent
experiment in the $\nu=1/3$ FQHL \cite{r6} ( $\nu$ is the
filling of the Landau levels ).

In the case $\nu=2/3$, on the other hand, it has been proposed that
there are two edge modes for each edge
corresponding to the $\nu=1$ QHL and $\nu=-1/3$ FQHL \cite{r2,r7}.
In this case the two edge modes are not spatially
separated and generally interact with each other.
Recently Kane et al. \cite{r8} studied this case and found that the
Hall conductance due to these edge modes is not quantized and
takes a nonuniversal value depending upon the coupling constant,
which contradicts the experiments observing the quantized Hall conductance
plateau at $\nu=2/3$.
The resolution of this puzzle they proposed is
that the randomness at the edge makes the neutral
mode massive and only the charged mode remains massless,
which gives the quantized Hall conductance.

Haldane \cite{r9} have opposed to their conclusion by showing
that the Hall conductance is quantized
without any randomness by taking the anomaly into account.
He also proposed the idea of topological (T-) stability of the
chiral edge modes, without which the quasi-particles with opposite
chiralities are generated by the randomness at the edge and the
FQHL becomes unstable.

If one regards the FQHE as the bulk phenomenon, however, it
should be insensitive to the details at the edge and the Hall
conductance should be always quantized.
In addition, other sets of experiments show the importance of the bulk
currents.  One example is the measurement of the Hall voltage in the sample,
which shows the existence of the voltage drop in the bulk and hence the
Hall current \cite{r10}. Another is the sample size dependence of the critical
current $I_c$ for the breakdown the QHE \cite{r11}. It is proportional to the
width of the sample, which shows that the current is distributed mainly in
the bulk.

In this paper we develop a theoretical
framework which treats the edge and bulk on an equal footing
in order to resolve the puzzles mentioned above.
Our theory is based on the Chern-Simons effective theory of FQHL
\cite{r12,r13,r14,r15}.
 Following the hierarchy construction by Jain \cite{r16} and by
 Blok and Wen \cite{r13},
 we decompose each of the original electrons into fictitious fermions.
Each fermion is represented as a composite particle of a boson
and a Chern-Simons gauge flux.
The external magnetic field $B_0$ is cancelled with the Chern-Simons gauge
flux on average, and the bosons are condensed to superfluidity \cite{r12}.
This condensation can be regarded as the hidden off-diagonal long
range ordering (ODLRO).
According to this picture the FQHE is the bulk phenomenon similar to the
superconductivity, and the edge current is analogous to the supercurrent
which is localized near the surface of the sample \cite{r17}.
The conserved current density $J_{I}^{\mu}$ of the bose condensate is
expressed in terms of the gauge field $a_{I \mu}$
($I=1,2, \cdot \cdot \cdot ,m$) in two spatial dimensions as
$J^{\mu}_I = { 1 \over 2\pi} \varepsilon^{\mu \nu \lambda}
\partial_{\nu} a_{I \lambda}
$.
A quasi-particle is represented as a vortex in the bose condensate,
which is defined as a zero point of the amplitude
of the boson order parameter. Thus the phase of the
boson order parameter there is singular.
The quasi-particle current density $j_I^{\mu}$ is defined as
$j_I^0(r) = \sum_{\ell} \delta(r -R_{I \ell}(t))$, and
$j_I^{\alpha}(r) = \sum_{\ell} { {d R_{I \ell}(t)} \over {dt}}
\delta(r -R_{I \ell}(t))$ $(\alpha=x,y)$,
where $R_{I \ell}(t)$ is the center of the $\ell$th vortex for the
$I$th bose condensate.
The effective Lagrangian for the Abelian FQHL is written in terms of the
gauge fields and the quasi-particle currents as \cite{r12,r13,r14,r15}
\begin{equation}
L = \int_S d^2 r \sum_{I}  \biggl[
\sum_J { 1 \over {4\pi} } K_{IJ} \varepsilon^{\mu \nu \lambda}
a_{I \mu} \partial_{\nu} a_{J \lambda} - a_{I \mu} j_I^{\mu}
- { 1 \over {2\pi}} A_{\mu}
\varepsilon^{\mu \nu \lambda}
\partial_{\nu} a_{I \lambda}
- { 1 \over {2\pi}} V(r)
\varepsilon^{\alpha \beta}\partial_{\alpha} a_{I \beta}
- { 1 \over g_I} f_{I \mu \nu} f^{\mu \nu}_I \biggr],
\end{equation}
where $K_{IJ} =K_{JI}$ is the $IJ$ component of the integer-valued
symmetric matrix $K$
representing the coupling between the $I$th and $J$th
bose condensates which uniquely specifies the topological structure of
the Abelian FQHL. It also gives the filling by
$\nu =\sum_{IJ} (K^{-1})_{IJ}$ ( see {\it  e.g.} \cite{r15}).
$V(r)$ is an arbitrary potential for the electrons.
The Maxwell term $ ( 1/ g_I ) f_{I \mu \nu} f^{\mu \nu}_I $
($ f_{I \mu \nu} = \partial_{\mu} a_{I \nu} - \partial_{\nu} a_{I \mu}$ )
in (1) is explicitly written as
$ ( 2 / g_I) [c_I^2 f_{I xy}^2 - f_{I0x}^2 - f_{I0y}^2]$, and
naturally arises in the duality mapping \cite{r12}.
The coupling constant $g_I$ is
given by $16\pi^2 \rho_{Is}/m_I$ where $\rho_{Is}$ is the
superfluidity density and $m_I$ is the mass of the bosons.
The velocity $c_I$ of the Bogoliubov mode for the $I$th
bose condensate is given by $c_I^2 = \rho_{Is}V_I/(2 m_I)$
where $V_I$ is the short-range repulsive interaction between the bosons.
The hard-core condition is realized by the effective $V_I$ with
$m_I V_I \sim 1$ in low energies.
The vector potential $A_{\mu}$ of the electromagnetic field
is coupled to the $\mu$ component of the physical current density
$J^{\mu} = \sum_I J^{\mu}_I$. Note that the vector potential
for the constant external magnetic field $ B_0$
has been already taken into account in the structure of the $K$ matrix, and
is not included in $A_{\mu}$. Similarly $a_{I \mu}$ and the density $J^0_I$
are measured from their average values in the following discussion.

The integral is over the sample $S$, and on the boundary $\partial S$
we impose
\begin{equation}
\sum_{\alpha = x, y} J_I^{\alpha} n_{\alpha} |_{\partial S}= 0,
\end{equation}
where $\vec n = ( n_x, n_y)$ is the unit vector normal to the boundary.
This boundary condition simply expresses the physical condition
that the current can not flow through the boundary $\partial S$.
Since it is a physical requirement, it is obviously invariant
with respect to a gauge
transformation $a_{I \mu} \to a_{I \mu} + \partial_\mu \phi_I$.
A remarkable fact is that the Chern-Simons term in the Lagrangian (1) is
also gauge invariant with the boundary condition (2).

We now derive the equation of motion by requiring that the variation
of the action $A = \int dt L$ vanishes.
The result is
$ \sum_J K_{IJ} \varepsilon^{\mu \nu \lambda} \partial_{\nu} a_{J \lambda}
= 2 \pi j_I^{\mu} +
\varepsilon^{\mu \nu \lambda} \partial_{\nu} (A_{\lambda}+
\delta_{\lambda 0} V)+ 8\pi \partial_{\nu} f^{\mu \nu}_I/g_I$
which is expressed in terms of the current density only and gauge invariant.
It is written
\begin{equation}
\left[
\begin{array}{ccc}
K & - 8\pi g^{-1} \partial_t & - 8 \pi c^2 g^{-1} \partial_{y} \\
8\pi g^{-1} \partial_{t}& K & 8\pi c^2 g^{-1} \partial_{x} \\
- 8\pi g^{-1} \partial_y & 8 \pi g^{-1} \partial_x & K
\end{array} \right]
\left[
\begin{array}{c}
{\bf J}^x \\
{\bf J}^y \\
{\bf J}^0 \\
\end{array} \right]
={ 1 \over {2 \pi}}
\left[
\begin{array}{c}
-(E_y+ \partial_y V) {\bf q} \\
(E_x + \partial_x V) {\bf q} \\
 B {\bf q} \\
\end{array}
\right]
+
\left[
\begin{array}{c}
{\bf j}^x \\
{\bf j}^y \\
{\bf j}^0 \\
\end{array} \right]
 \end{equation}
where $K=\{ K_{IJ} \}$, $g^{-1} =
diag( g_1^{-1}, \cdot \cdot , g_m^{-1} )$,
and $c^2 = diag( c_1^2, \cdot \cdot , c_m^2 )$ are $m \times m$ matrices
and
${\bf J^x} = ^t[J_1^x, \cdot \cdot , J_m^x]$,
${\bf j^x} = ^t[j_1^x, \cdot \cdot , j_m^x]$,
${\bf q} = ^t[1, \cdot \cdot , 1] $,
etc. are  vectors with $m$ components. Here we have introduced
the "charge vector" ${\bf q}$
representing the coupling to the external electromagnetic field \cite{r9}.
Equations (2) and (3) together with the Maxwell equations for
$A_{\mu}$ constitute the fundamental equations, from which all the
results below are obtained.
In the following sections 1 and 2
we shall neglect quasi-particles, i.e., vortices,
and set $j_I^{\mu}$ to be zero.
In section 3 the effects of the quasi-particles will be considered.

\noindent
{\bf 1. Quantization of the Hall Conductance}

 In order to study the quantization  it is enough to
consider the stationary case
$\partial_t J_I^{\mu} = \partial_t A_{\mu} =0$. Then (3) becomes
$K{\bf J^x} - 8 \pi c^2 g^{-1} \partial_y {\bf J^0} =
- (E_y+\partial_y V){\bf q}/(2\pi)$
and
$K{\bf J^y} + 8 \pi c^2 g^{-1} \partial_x {\bf J^0} =
(E_x+\partial_x V) {\bf q}/(2\pi)$.
Define the charging energy
$U(J^0_1, \cdot \cdot, J^0_m) \equiv
\int d^2 r \sum_I 2 f_I^{xy 2}/g_I =
\int d^2 r \sum_I
8 \pi^2 c^2_I (J^0_I)^2/g_I $, which is present in the Maxwell term of the
Lagrangian (1).
Then the physical current $J^{\alpha} = \sum_I J_I^{\alpha}$ is given by
\begin{equation}
J^{\alpha}(r) = - \varepsilon^{\alpha \beta} \partial_{\beta} R(r),
\end{equation}
where $R(r)$ is defined by
\begin{equation}
R(r) \equiv  { 1 \over {2 \pi}} \biggl[ \sum_{IJ} (K^{-1})_{IJ}
(A_0(r)+V(r))
- \sum_{IJ} (K^{-1})_{IJ} { {\delta U} \over {\delta J^0_J(r)} } \biggr].
\end{equation}
Let us now integrate (4) from $A$ to $B$ along the
contour $C$ in Fig.1. The result is that the
total current $I$ flowing across the contour $C$ is
$I = \Delta R$
where $\Delta R= R(B) - R(A)$.
The Hall conductance $\sigma_H$ is given by $I=\sigma_H \Delta \mu$,
where $\Delta \mu = \mu(B) - \mu(A)$ and
$\mu(r) = A_0(r) + V(r) - \delta E(\{ J^0(r) \})/ \delta J^0(r)$ is the
chemical potential. If $\Delta R$ is proportional to $\Delta \mu$,
, the Hall conductance is given by their ratio
$\sigma_H = \Delta R/\Delta \mu$.
In fact we shall show that this is the case if a condition on $K$ matrix is
satisfied.
The physically realized charging energy $E(J^0)$ is obtained by minimizing
the function
$U(J^0_1, \cdot \cdot, J^0_m)$
with the constraint
$\sum_I J^0_I = J^0$. Then it can be shown that \cite{com1}
\begin{equation}
{ { \delta E(J^0)} \over {\delta J^0} } = { 1 \over m} \sum_I
{{ \delta U(J^0_1, \cdot \cdot, J^0_m) } \over { \delta J^0_I} }
\end{equation}
for quite general functional $U$.
The comparison of (5) and (6) leads that $R(r)$ is proportional to
$\mu(r)$ if and only if the condition
\begin{equation}
\sum_I (K^{-1})_{IJ} = { 1 \over m}\sum_{IJ} (K^{-1})_{IJ}
= { \nu \over m}
\end{equation}
is satisfied, and $\sigma_H = \nu/2\pi$.
This is a striking result of the effective theory in the following respects.

\noindent
(i) $\sigma_H$ is always quantized for arbitrary shape of the sample
and potential $V(r)$ as long as the quasi-particle current $j_I^{\mu}$ is
absent.
In Fig.1 we can move the point $B$ (or $A$) freely as long as
it does not cross the current terminal, and the integral of (4) along $C$
remains the same. This concludes that the
the chemical potential $\mu(r)$ remains the same along the edge.
The chemical potential drop $\Delta \mu$ occurs
only at the current terminals.
Similarly the configuration of the 4 terminals $D$,$E$,$F$,and $G$
at one of the edge in Fig.1 is topologically almost equivalent to that
in the above discussion.
As before we integrate (4) along the path $C'$ in the bulk connecting
$D$ and $E$,  and we obtain the
quantized Hall conductance $\sigma_{xy} = I/\Delta \mu = \nu /2\pi$ \cite{r9}.

\noindent
(ii) The relation $\sigma_H =\Delta R/\Delta \mu = \nu/2 \pi$ is derived
without assuming the infinitesimal $I$ or $\Delta \mu$.
Therefore we believe that our theory goes beyond the linear response theory,
and remains valid for current $I$ less than the critical value $I_c$ for
the breakdown phenomenon.

\noindent
(iii) In the integral of (4) along the path $C$ or $C'$ in Fig.1, the bulk
current contributions can not be neglected as shown below.
Therefore the quantization of $\sigma_H$ occurs only when
both the bulk and edge currents are treated in a self-consistent way.
Our picture is that the quantization is the bulk phenomenon like the
superconductivity \cite{r12},
and hence robust against the perturbations at the edge as shown below.

The condition (7) is equivalent to a condition that the
"charge vector" ${\bf q}$ is an
eigenvector of the $K$ matrix in the notation of Ref.\cite{r9}.
The condition (7) is satisfied for the type
$K = I + p P$ where $I$ is the unit matrix, $p$ is an even integer,
and $P$ is the matrix with all the matrix elements being 1.
This $K$ matrix describes
the FQHL with $\nu = m/(1 + mp)$ including $\nu = 2/3, 2/5,$ etc
\cite{r13,r14,r15}.
Obviously this condition (7) is not satisfied for general
$K$ matrices and gives a criterion for the realization of
each Abelian quantum Hall state.

\noindent
{\bf 2. Edge Effects and Anomaly}

 The above discussion does not explicitly distinguish between the
bulk and edge effects. Now we concentrate on the edge properties.

\noindent
{\it Current and Charge Distribution near the Edge-}
For the steady state we set $\partial_t J_I^{\mu} = \partial_t A_{\mu}=0$
in (3).
Consider the sample extending the semi-infinite
plane ($x<0$) with the straight edge at $x=0$ and $V(r) =0$.
Since there is no spatial dependence in
the $y$-direction, $J_I^{x}=E_y=0$ in the stationary state.
For simplicity let us consider the case where
$\nu = 1/(2 n+1) \equiv \eta^{-1}
 = K^{-1}$ ($n$: integer). Only one gauge field is enough ( $m=1$) in this
case, and (3) is in the simplest form in which the $m \times m$ matrices
and vectors are c-numbers.
Then it becomes
$ J^0 - \kappa^{-2} d^2 J^0/ d x^2 = (8\pi / g \eta^2) d^2 A_0 /d x^2$,
where $A_0(x)$ is the scalar potential and related to the electric field
$E_x$ as $E_x = - d A_0/dx$.
$\kappa^{-1} \equiv 8 \pi c/g \eta$ is the characteristic length scale
of the spatial change.
The magnitude of $\kappa^{-1}$ is estimated as $\kappa^{-1} \sim \ell_B
\sim \sqrt{ \hbar/e B_0}$ (magnetic length).
This together with
$A_0(x) = - 2e^2 \int d x' \ln | x - x'| J^0(x')$
constitutes self-consistent equations.
These equations without the term
$ - \kappa^{-2} d^2 J^0/ d x^2$
has been already studied by several authors \cite{r18}.
They obtained the charge and voltage drop
localized near the edge, but the localization length $W$ is of the
order of $W \sim \sqrt{L_x e^2/ \hbar \omega_c}$
with $\omega_c$ being the cyclotron frequency and $L_x$ is the
sample width.
In our case $L_x \to \infty$ and the current distribution has power-law tail
( $ \propto x^{-2}$ ) without any length scale.
For a typical sample size $W$ is much larger than the
magnetic length $\ell_B \sim \kappa^{-1}$, and
$- \kappa^{-2} d^2 J^0/ d x^2$ can be safely neglected and our analysis is
consistent with the previous ones \cite{r18}.
Thus the current distribution is not localized near the edge within the
length scale of $\ell_B$ where the edge mode is localized as shown
shortly.
Therefore the bulk current can never be neglected when one takes the Coulomb
interactions into account.

\noindent
{\it  Edge Mode as a Self-Induced Eigenmode of the FQHL} \cite{r17}-
We set $E_x=E_y=B = 0$ in (3).
 Let us first consider the case of $\nu = 1/(2 n+1)$ ($m=1$) and the
the semi-infinite sample ($x<0$) described above.
The edge mode is derived by assuming
$J_x=0$, $J^y = J^y_0 e^{\gamma x} e^{ -i \omega t + i k_y y}$, and
$J^0 = J^0_0 e^{\gamma x} e^{ -i \omega t + i k_y y}$.
Putting these into (3), we obtain two decoupled equations. One is
from the 1st row of (3) and is given as
\begin{equation}
\omega J^y_0 - c^2 k_y J^0_0 = 0.
\end{equation}
The other is from the 2nd and 3rd rows and is the eigenvalue problem for the
inverse of the penetration length $\gamma$ as
\begin{equation}
\left[
\begin{array}{cc}
\eta &  8 \pi c^2 g^{-1} \gamma \\
8\pi g^{-1} \gamma & \eta
\end{array} \right]
\left[
\begin{array}{c}
J_0^y \\
J_0^0 \\
\end{array} \right]
=
\left[
\begin{array}{c}
0 \\
0 \\
\end{array}
\right],
\end{equation}
which gives $\gamma = \pm \kappa = \pm g \eta/8 \pi c$
and $-J^y_0 = \pm c J^0_0$.
Here we take the solution with positive $\gamma$ in order that it is
non-diverging and localized near the edge.
Putting this solution into (8) we obtain the dispersion
relation of this mode $ \omega = - c k_y$, where the velocity $c$ is that of
the Bogoliubov mode for the bose condensate.
Hence we have obtained  the chiral edge mode with the linear
dispersion relation microscopically from (2) and (3),
which has been assumed previously \cite{r2}.
This derivation can be generalized to the hierarchy cases ($m>1$)
in a straightforward way, but the calculation is rather
complicated. The electromagnetic responses, however, can be analyzed
without solving the eigenvalue problem as will be discussed in the
following section.

\noindent
{\it Electromagnetic Responses of the Edge Modes-} We
calculate the response $J^{\mu}(x,k_y,\omega)$ to the
dynamical external field $A_{\mu}$ with the frequency
$\omega$ and the wavenumber $k_y$ along the edge.
We first study the simplest case of $\nu=1/(2n+1)$ ($m=1$), and
later the hierarchy cases ($m>1$).
Then (3) with the boundary condition (2) can be
solved as
\begin{eqnarray}
J^0(x, k_y,\omega) &=& { 1 \over { 2\pi \eta}} \kappa e^{ \kappa x}
 { {i E_y(x=0, k_y,\omega)} \over { k_y + \omega/c} }  +
 {\rm terms \ \ containing} \ \ E_x, B
\nonumber \\
J^y(x, k_y,\omega) &=& -{ 1 \over { 2\pi \eta}} \kappa e^{ \kappa x}
 { {i E_y(x=0, k_y,\omega)} \over { k_y + \omega/c} }  +
 {\rm terms  \ \ containing} \ \ E_x, B.
\end{eqnarray}
We interpret the term  containing $E_y(x=0)$ as the charge and current
of the edge channel because they are localized near the edge and
also proportional to the electric field along the edge \cite{r3}.
Therefore the edge channel contribution was not included in the
discussion above where $E_y$ was set to be zero.
The terms containing $E_x$ and $B$ are regarded as the bulk current
and charge.
The total edge current $I^y_{\rm edge}$ and the edge charge
$I^0_{\rm edge}$ are the integrals of the above edge contributions
over $x$. We obtain
$I^y_{\rm edge} = -(\nu/2 \pi) i E_y(x=0)/(k_y - \omega/c)$,
which implies that the conductance of the edge channel is $\nu/2 \pi$
when considering the case  $\omega=0$ and $E_y = - \partial_y A_0
= - i k_y A_0$.

For the hierarchy cases ($m>1$), it can also be shown that  the
edge channel contributions can be singled out by setting $E_x=B=0$ and
$E_y(x=0) \ne 0$ as in the case of $\nu=1/(2n+1)$ ($m=1$).
Then we integrate the second and third row of (3) along the $x$ direction
to obtain
$ K {\bf I_{\rm edge}^y} \equiv K \int^0 dx {\bf J^y} = - 8 \pi c^2 g^{-1} {\bf
J^0}(x=0)$
and
$K {\bf I_{\rm edge}^0} \equiv K \int^0 dx {\bf J^0} =
- 8 \pi g^{-1} {\bf J^y}(x=0)$.
Putting these into the first row of (3) at $x=0$, we obtain the
anomaly equation $
K (\partial_t {\bf I_{\rm edge}^0} + \partial_y {\bf I_{\rm edge}^y} )
= E^y(x=0){\bf q}/2\pi $.
For the physical edge charge $I_{\rm edge}^0$ and current $I_{\rm edge}^y$
the anomaly equation is \cite{r9}
\begin{equation}
\partial_t I_{\rm edge}^0 + \partial_y I_{\rm edge}^y = (\nu/2\pi) E_y,
\end{equation}
which implies that the edge conductance $\sigma_{\rm edge}$
is $\nu/2\pi$ because this equation becomes
$\partial_y( I_{\rm edge}^y + (\nu/2\pi)A_0 ) = 0$
when one considers the stationary flow.
This $\sigma_{\rm edge}$ is, however, different from the physical Hall
conductance $\sigma_H$ due to the following reasons:
(i)The edge is the equi-potential line in the steady state, and
the voltage drop occurs at the current terminals as shown before, which
are not taken into account in (10) and need special treatment \cite{r9}.
(ii) Even though the current terminals are taken into account, the Coulomb
repulsion spreads the current distribution much deeper into the sample than
the penetration depth $\kappa^{-1} \sim \ell_B$ of the edge mode.
(iii) The voltage drop is not the same as the chemical potential difference
since the charging energy is present.

$\sigma_{\rm edge}$ is not measured by the Hall conductance but
rather by some optical experiments \cite{r2}.
Then it is remarkable that $\sigma_{\rm edge} =\sigma_H$ when $V(r)=0$.
$\sigma_{\rm edge}$ is susceptible to the randomness at the edge as
stressed by Kane et al. \cite{r8}. $\sigma_H$, on the other hand, is robust
against random potential $V(r)$ if $j^{\mu}_I =0$.
Thus it is possible that  the values of
$\sigma_{\rm edge}$ and $\sigma_H$ are different.

\noindent
{\bf 3. Quasi-Particles}

We now consider the effects of the
quasi-particles which can not be neglected  when we
take into account a random potential at the edge.
In the presence of the nonzero quasi-particle (vortex) current
$j^{\mu}_I$ in the r.h.s. of
(3), the quantization of the Hall conductance breaks down.
Thanks to the Maxwell term the
spatial dependence of the fields are smooth even near the edge,
and the only possible singularity is that of the vortex (quasi-particle).
Therefore it is easy to distinguish between the current due to the
quasi-particle and the bose condensate.
 According to this criterion the so called "quasi-particle" in the
edge channel is represented as the kink of the condensate field variables
and have no singularity, and hence does not contribute to the
quasi-particle current density $j^{\mu}_I$.
This can be viewed in the following way.
Consider a vortex inside the sample. This has the finite energy
in the FQHL at plateau, and have accordingly tends to be excluded to
the outside of the sample like a magnetic flux in the
superconductor. When the center of the vortex $R_{I\ell}$
goes out of the sample \cite{com2},
it will leave a smooth phase change nearby, which can be identified as
a kink in the edge mode.
With backward scatterings by the randomness near the edge,
pair creations of the vortices at different edge modes occurs.
The center of the vortices, however, will be repelled from the sample.
These processes have been described in terms of the charged
operators like $e^{i \phi}$ where $\phi$ is the bose field
for the chiral edge mode \cite{r8}.
In this case the quasi-particle current $j^{\mu}_I$ is localized near the
random potential. This does not mean the effects of
the randomness is also localized. It is possible that the
edge modes becomes massive. If we can make the path where $j^{\mu}_I =0$
connecting the two voltage terminals, however, the discussion above
is not modified. The Hall conductance remains to be quantized even if the
edge modes becomes massive.
In the case of the point contact which has been studied in \cite{r6},
on the other hand, a path of the integral (4) always crosses
the region where the
quasi-particle current $j^{\mu}$ is nonzero. Therefore
the quantization of the
Hall conductance breaks down and we expect the voltage drop across the
point contact where the quasi-particle tunneling between the
two edges of the sample occurs.

 In summary, we have proposed an effective theory of fractional quantum
Hall liquid with edges in the dual representation,
which describes the edge and bulk in a unified way. The Maxwell term
cures the pathology of the topological theory which has no Hamiltonian.
The edge modes with anomaly are derived, and the distribution of the
current, charge, and Hall voltage can be also calculated.
With the condition $\sum_I (K^{-1})_{IJ} = \nu/m$ and no quasi-particle in the
bulk, we obtain the quantization of the Hall conductance irrespective of the
detailed dynamics and the randomness at the edge.

\acknowledgements
We are grateful to T.K.Ng, A.Zee, X.G.Wen, and Y.S. Wu
for useful discussions. This work is supported by Grant-in-Aid for
Scientific Research No. 04240103
from the Ministry of Education, Science, and Culture of Japan.

\figure{ Hall bar with voltage terminals at $A$,$B$ and $D$,$E$, and
the current terminals at $F$,$G$.
$I$ is the total current flowing in and out while $J$ is the
current density in the sample.}


\begin{references}

\bibitem{r1}F.D.M. Haldane, J.Phys. C{\bf{14}}, 2585 (1981).

\bibitem{r2}X.G. Wen, Phys. Rev. B{\bf{41}}, 12838 (1990);
Phys. Rev. B{\bf{43}}, 11025 (1991);
Phys. Rev. Lett. {\bf{64}}, 2206 (1990);
D.H.Lee and X.G.Wen, Phys. Rev. Lett. {\bf{66}}, 1765 (1991).

\bibitem{r3}M. Buttiker, Phys. Rev. B{\bf{38}}, 9375 (1988).

\bibitem{r4}C.L. Kane and M.P.A. Fisher, Phys. Rev. Lett. {\bf{68}}, 1220
(1992);
A. Furusaki and N. Nagaosa, Phys. Rev. B{\bf{47}}, 3827 (1993).

\bibitem{r5}K. Moon, H. Yi, C.L. Kane, S.M. Girvin, and M.P.A. Fisher,
Phys. Rev. Lett. {\bf{71}}, 4381 (1993).

\bibitem{r6}F.P. Milliken, C.P. Umbach, and R.A. Webb, unpublished.

\bibitem{r7}A.H. MacDonald, Phys. Rev. Lett. {\bf{72}}, 220 (1990).

\bibitem{r8}C.L. Kane, M.P.A. Fisher, and J. Polchinski,
Phys. Rev. Lett. {\bf{72}}, 4129 (1994).

\bibitem{r9}F.D.M. Haldane, Phys. Rev. Lett. {\bf{74}}, 2090 (1995).

\bibitem{r10}H.Z. Zheng, D.C. Tsui, and A.M. Chang,
Phys. Rev. B{\bf{32}}, 5506 (1985).

\bibitem{r11}S. Kawaji et al., J.Phys. Soc. Jpn. {\bf{63}}, 2303 (1994).

\bibitem{r12}S.C. Zhang, Int. J. Mod. Phys. B{\bf{6}}, 25 (1992)
and references therein.

\bibitem{r13}B. Blok and X.G. Wen,
Phys. Rev. B{\bf{42}}, 8133, {\it ibid.} 8145 (1990).

\bibitem{r14}J. Frohlich and A. Zee,
Nucl. Phys. B{\bf{364}}, 517 (1991).

\bibitem{r15}X.G. Wen and A. Zee,
Phys. Rev. B{\bf{46}}, 2290 (1992).

\bibitem{r16}J.K. Jain, Phys. Rev. B{\bf{40}}, 8079 (1989).

\bibitem{r17}N. Nagaosa and M. Kohmoto, in
{\it Correlation Effects in Low-Dimensional Electron
Systems}, eds. A.Okiji and N.Kawakami (Springer-Verlag, 1994) p168.

\bibitem{com1}The details of the derivation will be published elsewhere.

\bibitem{r18}A.H. MacDonald, T.M. Rice, and W.F. Brinkman,
Phys. Rev. B{\bf{28}}, 3648 (1983); D.J.Thouless,
J. Phys. C{\bf{18}}, 6211 (1985).

\bibitem{com2}Of course the vortex outside of the sample is a fictitious
object expressing the phase configuration inside the sample.
\end{references}
\end{document}